
\magnification=1200
\hsize 16.5truecm
\vsize 22truecm

\font\footfont=cmr8

\fontdimen16\tensy=2.7pt
\fontdimen17\tensy=2.7pt
\fontdimen14\tensy=5pt

\countdef\refno=30
\refno=0
\countdef\sectno=31
\sectno=0
\countdef\chapno=32
\chapno=0
\newskip\zmineskip \zmineskip=0pt plus0pt minus0pt
\mathchardef\mineMM=20000
\newinsert\footins
\def\footnote#1{\footfont\let\minesf\empty 
  \ifhmode\edef\minesf{\spacefactor\the\spacefactor}\/\fi
   #1\minesf\vfootnote{#1}}
\def\vfootnote#1{\insert\footins\bgroup
  \interlinepenalty\interfootnotelinepenalty
  \splittopskip\ht\strutbox 
  \splitmaxdepth\dp\strutbox \floatingpenalty\mineMM
  \leftskip\zmineskip \rightskip\zmineskip
\spaceskip\zmineskip \xspaceskip\zmineskip
 \item{#1}\footstrut\futurelet\next\fominet}
\def\fominet{\ifcat\bgroup\noexpand\next \let\next\fmineminet
  \else\let\next\fminet\fi \next}
\def\fmineminet{\bgroup\aftergroup\minefoot\let\next}
\def\fminet#1{#1\minefoot}
\def\minefoot{\strut\egroup}
\def\footstrut{\vbox to\splittopskip{}}
\skip\footins=\bigskipamount 
\count\footins=1000 
\dimen\footins=8in 
\def\ref{\advance \refno by 1 \ifnum\refno<10 \item{ [\the\refno ]} \else
\item{[\the\refno ]} \fi}
\outer\def\section#1#2\par
           {
           \vskip0pt plus .3\vsize\penalty-250\vskip 0pt plus-.3\vsize
           \bigskip\vskip\parskip
           \noindent\leftline{\rlap{\bf #1}
           \hskip 17pt{\bf #2}}
            \nobreak\smallskip}
						
\def\eq{\the\@eqtoks}\def\eqn{\the\@eqnotoks}%
\def\@ldisplaytest#1\leqno#2\leqno#3\@ldisplaytest{%
\@eqtoks = {#1}%
\if #3%
\eqnofalse
\else
\eqnotrue \leqnotrue
\@eqnotoks = {#2}%
\fi}%

\def\generaldisplay{%
\leftline{%
\strut \indent \hskip\leftskip
\ifeqno \ifleqno
\rlap{$\displaystyle\eqn$}%
\kern\dimen0
\fi\fi
$\displaystyle{\eq}$%
\ifeqno  \ifleqno\else
\hfill $\displaystyle{\eqn}$%
\fi\fi}}%
\def\centereddisplays{\let\displaysetup = \relax}

\newif\ifeqno \newif\ifleqno \everydisplay{\displaysetup}
\def\displaysetup#1$${\displaytest#1\eqno\eqno\displaytest}
\def\displaytest#1\eqno#2\eqno#3\displaytest{%
\if!#3!\ldisplaytest#1\leqno\leqno\ldisplaytest
\else\eqnotrue\leqnofalse\def\eqn{#2}\def\eq{#1}\fi
\generaldisplay$$}
\def\ldisplaytest#1\leqno#2\leqno#3\ldisplaytest{\def\eq{#1}%
\if!#3!\eqnofalse\else\eqnotrue\leqnotrue\def\eqn{#2}\fi}

\def\rto{ \lower1.7ex\vbox{
\hbox{$ {\hbox to 3cm{ \rightarrowfill } } $ }\lineskip=0pt\baselineskip=0pt
\hbox{\kern2.7em
{$
{\scriptstyle{R \to \infty }}  $}  }}}

\def\real{{\vrule height 1.5ex width 0.05em depth 0ex
\kern -0.03em {\rm R}}}
\def\ostar{\hbox{\lower0.4ex\hbox{*} \kern -2.4ex \hbox{\sl O}} }
\def\oostar{\leavevmode {\kern 0.15em\lower0.5ex
\hbox{*}}\leavevmode \kern -0.65em \raise-0.12ex  \hbox{O}}
\def\sstar{\hbox{\lower0.8ex\hbox{*}}}

\baselineskip=12pt
\null
\hfuzz=15pt
\vskip 5.5cm
\centerline{\bf QUANTUM $\kappa$-POINCARE IN ANY DIMENSION}
\vskip 1.5cm
\centerline{\bf Jerzy LUKIERSKI \footnote{$^{1)}$}{On leave of absence
from the Institute for Theoretical Physics, University of Wroclaw,
pl. Maxa Borna 9, 50-204 Wroclaw, Poland.}
\footnote{$^{2)}$}{Partially supported by the Swiss National Science
Foundation.}
\footnote{$^{3)}$}{Partially supported by KBN grant Nr 2/0124/91/01.}
{\bf and Henri RUEGG$^{2)}$}}
\medskip
\centerline{D'partement de Physique Th'orique}

\centerline{Universit' de GenŠve}

\centerline{24, quai Ernest-Ansermet}

\centerline{CH-1211 GenŠve 4 - Switzerland}
\vskip 2cm
\centerline{\bf Abstract}
\bigskip
The $\kappa$-deformation of the D-dimensional Poincar\'e algebra $(D\geq
2)$ with any signature is given. Further the quadratic Poisson brackets,
determined by the classical $r$-matrix are calculated, and the quantum
Poincar\'e group "with noncommuting parameters" is obtained.
\vfill\eject
\noindent
{\bf 1. Introduction}
\medskip
The quantum deformation of semisimple Lie algebras has been given by
Drinfeld [1] and Jimbo [2]. However such a general solution does not
exist for non
semisimple Lie algebras. Examples of quantum inhomogeneous rotation
algebras
with Euclidean or Minkowski metric have been given for the dimensions two
[3]
three [4] and four [4,5], and shown to be Hopf algebras.
The
aim of this paper is to point out the  common features of these examples
and to generalize them to any dimension.

The method used in these examples is the contraction procedure, applied
to the Drinfeld-Jimbo deformation of the semisimple rotation algebras in three,
four and five dimensions, denoted by $U_q (so(D)), D= 3,4,5.$
In this way we obtain the Hopf algebra $U_q (so(D-1) \supset
T_{D-1})$, i.e. the deformation of the semidirect sum of rotations and
translations in $D-1$ dimensions. The presence of the "de Sitter" radius R
(which is ultimately taken to infinity) explains that the deformation
parameter $q$ is replaced by a dimensionful parameter which we called
$\kappa$ [4,5] for $D-1=4$. For this reason we denoted the
quantum deformation of the enveloping algebra of the Poincar\'e algebra
$P_4$  in four dimension by
$U_{\kappa} (P_4)$. In the limit $\kappa \rightarrow \infty$  one recovers
$P_4$ from the Cartan-Weyl basis.

Another way to get $U_\kappa (P_4)$ is to start with standard $P_4$ and
replace the right hand side of the commutation relations involving the
boosts by functions of the momenta (with some restrictions to be
specified
later). The Jacobi identities restrict these functions in such a way that one
of
the solutions is
precisely $U_\kappa (P_4)$ [6,7]. For dimensional reasons, the
momenta appearing on the RHS of the commutator of two boosts must
be divided by a dimensionful constant which turns out to be the
deformation parameter $\kappa$.

One can show that it is not possible to write a quantum Poincar\'e
algebra as a semidirect sum of a quantum Lorentz algebra and a quantum
fourmomentum with standard (Drinfeld-Jimbo) deformation of the Lorentz
sector unless one adds the dilatation generator [8,9]. An explicit
example of such a quantum deformation of the Poincar\'e group as
well as Poincar\'e algebra with nonstandard deformation of the Lorentz
sector was given recently by Chaichian and Demichev
[10]. Together with the quantum $\kappa$-Poincar\'e algebra
[4,5], these are the only Poincar\'e Hopf algebras with 10
generators so far known.

{}From the point of view of physics it is
interesting
to compare the predictions of $U_\kappa (P_4)$ with those of the $P_4$.
The modification of g-2 [11] imposes a lower limit of $10^7$ Gev for
$\kappa$,
while the change in the quadratic Casimir operator requires $\kappa
>10^{12}$ Gev [12]. For other predictions see [13,14].

An interesting result was obtained by Maslanka [15]. He performed a non
linear change in the Cartan-Weyl basis of the classical Poincar\'e Lie
algebra. The new generators obtained in this way belong to $U(P_4)$ but
they obey the commutation relations of the generators of $U_\kappa
(P_4)$. From this one obtains the $\kappa$-Poincar\'e algebra with
(complicated)
cocommutative coproducts. Such a
coproduct is actually preferred by some authors [6] because of the
difficulties of the physical interpretation of a non cocommutative
coproduct (for dimensions of space-time larger than two).
In the present paper we shall assume however that the coproduct is
noncocommutative, i.e. we shall assume a genuine quantum deformation
of the Poincar\'e algebra.
\vfill\eject
\noindent
{\bf 2. Quantum $\kappa$-Poincar\'e in 4-dimensions}.
\medskip
In order to abstract the common features of the quantum deformations of
inhomogeneous rotation algebras we start with the explicit example of
$U_\kappa (P_4)$ for a real value of $\kappa$ [5]. We choose the
Minkowski metric (--+++), but it is easy to change the metric to the
Euclidean
(++++) or to (++-- --) [7].

The Hopf algebra structure is the following: (we use $\vec M$ for
rotations, $\vec N$ for boosts, $P_\mu$ for momenta).
\noindent
\medskip
\underbar{ Algebra structure:}
$$
\eqalign{[M_i,M_j] &= i\epsilon_{ijk} M_k \qquad    i,j, k=1,2,3\cr
[M_i, P_0]&=0\cr
[M_i, P_j]&=i\epsilon_{ ijk} P_k\cr
[M_i, N_j]&=i\epsilon_{ijk} N_k}
\eqno (2.1)
$$

$$
[P_\mu, P_v] =0 \qquad  \mu, v=0,...3
\eqno (2.2)
$$

$$
\eqalign{[N_i, P_0] &= iP_i\cr
[N_i, P_j] &= i\delta_{ij} \kappa \sinh {P_0 \over \kappa}}
\eqno (2.3)
$$

$$
\eqalign{[N_i, N_j] &= -i \epsilon_{ijk} M_k \cosh {P_0 \over {\kappa}}
+ i X_{ij}\cr
X_{ij} &= {1\over {4\kappa^2}} \epsilon _{ijk} P_k (\vec P \vec K)}
\eqno (2.4)
$$

It is seen that the commutation relations involving the rotations $M_i$
and
the time translation $P_0$ are classical, and the momenta commute. The
quantum
deformation appears only when the boosts are involved. For dimension $D=3$
the
classical $so(3)$ is replaced by $so(2)$.

For $D=2$ there is only one boost, and (2.4) is trivial, while (2.3) is
the same for $D=2,3,4$.

For $D=3$, (2.4) is linear in $M_3$ cosh $P_0{/\kappa}$, but the term
$X_{ij}$ linear in
$\vec M$ and bilinear in $\vec P$ is absent, because there are only two
space momenta and one rotation $M_3$.

We now discuss the
\noindent
\medskip
\underbar {Coalgebra structure}

$$
\eqalign {\Delta (M_i)&= M_i\otimes I + I \otimes M_i\cr
\Delta (P_0)&=P_0\otimes I + I\otimes P_0}
\eqno (2.5)
$$

$$
\Delta (P_i)= P_i\otimes e ^{P_0/2\kappa}+ e^{-P_0/2 \kappa}
\otimes P_i
\eqno (2.6)
$$

$$
\eqalign{\Delta(N_i)&=Ni\otimes e^{P_0/2\kappa} + e
^{P_0/2\kappa}\otimes N_i + Y_i\cr
Y_i&={i\over 2\kappa} \epsilon_{ijk} \left(P_j \otimes M_k e ^{P_0/2\kappa}
+ e ^{-P_0/2\kappa} M_j\otimes P_k\right)}
\eqno (2.7)
$$

The coproducts for $M_i$ and $P_0$ are classical. For $D=2$, $Y_i$ is
absent

\noindent
\medskip
\underbar{Counits}
$$
\epsilon (M_i) = \epsilon (P_{\mu}) = \epsilon (N_i)= 0
\eqno (2.8)
$$

\noindent
\underbar {Antipodes}
$$
\eqalign{S (M_i)& = M_i\cr
S (P_{\mu}) & = P_{\mu}\cr
S(N_i)& = - N_i + {3 i \over 2\kappa} P_i}
\eqno (2.9)
$$

\noindent
All the axioms of a Hopf algebra are satisfied: associativity of the
product
m defined on the algebra, coassociativity of the coproduct $\Delta$ of the
coalgebra; $\Delta$ and the counit $\epsilon$ are homomorphisms of the
algebra.
The following relations are satisfied.
$$
m\circ(S\otimes id)\circ\Delta = m\circ(id\otimes S)\circ\Delta =
i\circ\epsilon
\eqno (2.10)
$$
\noindent
where $i$ is the unit  map of the algebra.

For $D=2$, the term linear in $P_i$ for the antipode $S(N_i)$ is missing.

\bigskip
\noindent
{\bf 3. Quantum $\kappa$-Poincar\'e in any dimension.}
\medskip
We have seen that : \hfill\break
\medskip
\noindent
i) for $D=3,4$, the rotation subalgebra $o(D-1)$ remains
classical,
\medskip
\noindent
ii) for $D=2,3,4$ the translations commute,
\medskip
\noindent
iii) the time translation $(P_0)$
is classical, while the other(s) have a non trivial coproduct,
\medskip
\noindent
iv) equation
(2.3)
is the same in all three cases.

We shall keep these features for the quantum deformation of $o(D)$
$\supset T_D \equiv i(o(D)$), the inhomogeneous rotation algebra. We
require the "rotation" subalgebra o(D-1) to be classical, as well as
the
"time" translation $P_0$. The "space" translations $P_a$, $a=1..D-1$,
while commuting with $P_0$ and among themselves, will have a non trivial
coproduct.

The metric tensor  $g_{AB}, A,B=0..D-1$, will be kept general (diagonal).
We shall call "rotations" the generators $M_{ab} =- M_{ba}$,
$a,b,=1...D-1$,
and "boosts" the D-1 generators $M_{a0} = -M_{0a}$. The sign convention is
chosen in such a way that for the "Minkowski" metric $g_{AB}$ = diag
(-- ++...+), we recover for $D=4$ the $\kappa$-Poincar\'e algebra with
$M_{ij} = \epsilon_{ijk} M_k,\,\, i,j, k=1,2,3$ and $M_{i0} = N_i$. With
these notations we postulate the following
\bigskip
\noindent
\underbar{algebra structure}

$$
\eqalign{[M_{ab}, M_{cd}] &= -i (g_{ad} M_{bc} + g_{bc}
M_{ad}-g_{ac}M_{bd} -g_{bd}M_{ac})\cr
[M_{ab}, P_0] &=0\cr
[M_{ab}, P_c] &=-i(g_{bc}P_a-g_{ac}P_b)\cr
[M_{ab}, M_{c0}] &= -i(g_{ac}M_{a0} - g_{ac}M_{a0} )}
\eqno (3.1)
$$
$$
a,b,c,d = 1, \cdots ,D-1
$$

$$
[P_A, P_B] = 0 \qquad A,B = 0,1 ,\cdots , D-1
\eqno (3.2)
$$

$$
\eqalign{[M_{a0}, P_0] &= iP_a\cr
[M_{a0}, P_b] &=ig_{ab} \kappa \sinh {P_0 \over \kappa}}
\eqno(3.3)
$$

$$
[M_{a0}, M_{b0}] = i g_{00} M_{ab} \cosh {P_0\over \kappa} + i X_{ab}
\eqno (3.4)
$$
$X_{ab}$ has to satisfy the Jacobi identities.
This implies
$$
[P_a, X_{ab}] =0 \qquad a\not= b\ ,\ a\ {\rm fixed}
\eqno (3.5)
$$

$$
\eqalign{[M_{a0}, X_{bc}] &+ [M_{b0}, X_{ca}]+[M_{c0}, X_{ab}]=\cr
&= {i\over \kappa}  \sinh {P_0 \over \kappa} (M_{ab}P_c + M_{bc}P_a
+ M_{ca}P_b)}
\eqno (3.6)
$$

\noindent
To our surprise we found that $X_{ab}$ is linear in $M_{ab}$ and
\underbar{bilinear} in
$P_a$ for any dimensions! Thus the case D=4 is already the general
case.
The following satisfies the Jacobi identities:
$$
Y_{ab} = {1\over 4\kappa^2} P^d ( P_a M_{bd} + P_b M_{da} + P_d M_{ab})
\eqno (3.7)
$$
and reduces to (2.4) for $D=4$.

It is now easy to guess the
\medskip
\noindent
\underbar {Coalgebra structure}

$$
\eqalign{\Delta (M_{ab} &= M_{ab} \otimes I + I \otimes M_{ab}\cr
\Delta (P_0) &= P_0 \otimes I + I \otimes P_0}
\eqno (3.8)
$$

$$
\Delta (P_a) = P_a \otimes e^{P_0/2\kappa} + e^{-P_0/2\kappa}\otimes P_a
\eqno (3.9)
$$
$$
\Delta (M_{a0})= M_{a0} \otimes e^{P_0/2\kappa} + e^{-P_0/2\kappa}\otimes
M_{a0} + Y_a
\eqno (3.10)
$$

The expression for $Y_a$ is obtained from the homomorphism property of
the coproduct.

$$
Y_a = -{g_{00} \over 2\kappa} (P^b \otimes M_{ab} e^{P_0 /2\kappa}
+ e^{-P_0 /2 \kappa} M_{ab} \otimes P^b)
\eqno (3.11)
$$
which is of course the same as (2.7) for $g_{00}=-1$.

The counits are again :
$$
\epsilon(M_{AB}) = \epsilon(P_A)  = 0 \quad A,B = 0,1,...D-1\eqno (3.12)
$$
and the \underbar{antipodes} are calculated from equation (2.10).
$$
\eqalign{S(X) &=-X \qquad X =M_{ab}, P_A\cr
S(M_{a0}) &= -M_{a0} - {ig_{00}\over 2\kappa} [(D-2)g_{aa} + 1] P_a\ .}
\eqno (3.13)
$$
\bigskip
\noindent
{\bf 4. From Classical r-matrix to quantum Poincar\'e group}
\medskip
{}From the $\kappa$-Poincar\'e quantum algebra, with the coproducts given by
(3.8-10), one can obtain the classical r-matrix, i.e. one gets the
classical Poincar\'e bialgebra. Because such an r-matrix satisfies
the (modified) classical YB equation, one can construct the quadratic
Poisson brackets on the classical Poincar\'e  group $ISO(D))$ dual to the
Poincar\'e algebra. The quantization of these Poisson brackets provide the
quantum Poincar\'e group with non commuting parameters satisfying
quadratic relations.

Following Zakrzewski [16] who first provided the
quantization
scheme described above for $D=4$,
we define the cocommutator on the algebra $i(o(D))$ by the
antisymmetric
part of the coproducts (3.8-10) linear in ${1\over\kappa}\,\, (\tau (a
\otimes b)  \equiv b\otimes a)$
$$
\delta = \Delta - \tau \circ \Delta =\delta + 0 ({1\over\kappa^2})
\eqno (4.1)
$$
One obtains $(a \wedge b\equiv a\otimes b - b\otimes a)$

$$
\eqalign{\delta (M_{ac}) &= \delta (P_0) =0\cr
\delta (P_a) &= {1 \over \kappa} P_a \wedge P_0    \cr
\delta (M_{a0}) &= {1 \over \kappa} \{ M_{a0} \wedge P_0 - g_{00}
(P^b \wedge M_{ab}) \} }
\eqno (4.2)
$$

It appears that the cobrackets (4.2) can be expressed as follows
 $(X \epsilon  U(io(D)))$
$$
\delta (X) = [X \otimes 1+1\otimes X, r]
\eqno (4.3)
$$
where $r\, \epsilon\, U(io(D))\otimes U(io(D))$ is the classical
$r-$matrix for the Poincar\'e algebra.

$$
r =  \sum_{a=1}^{D-1} {-i\over \kappa} M_{a0} \wedge P^a
\eqno (4.4)
$$

The cocommutator (4.3) satisfies the Jacobi identity if and only if the
tensor defining the classical YB equation [1]
$$
[r,r ]_s := [r^{12}, r^{13}] + [r^{13}, r^{13}] + [r^{12}, r^{23}]
\eqno (4.5)
$$
is $io(D)$-invariant, i.e. $(X\epsilon U(io(D)))$
$$
[X \otimes 1 \otimes 1 + 1 \otimes X  \otimes 1 + \otimes 1 \otimes X,
[r, r]_s] = 0
\eqno (4.6)
$$
The expression (4.5) is equal to the Schouten bracket of a pair of
bivectors r [17]. For any Lie bialgebra $\hat g$ with the commutation
relations
$$
[e_\alpha, e_\beta]=f_{\alpha\beta}^\gamma e_\gamma
\eqno (4.7)
$$
and the cocommutator (4.3) with
$$
r  = r^{\alpha\beta}e_{\alpha} \wedge e_{\beta} \equiv r^{\alpha\beta}
(e_\alpha \otimes e_\beta - e_\beta \otimes e_\alpha)
\eqno (4.8)
$$
the Schouten bracket (4.5) takes the following explicit form :

$$
\eqalign
{[r, r]_s &=
 f^\alpha_{\mu v} r^{\mu\beta} r^{\nu\gamma} e_\alpha \wedge
e_\beta \wedge e_\gamma}
\eqno (4.9)
$$
Applied to ${\hat g}=o(D) \supset T_n \equiv i (o(D))$, this gives
$$
[r,r]_s = {i\over \kappa^2} (g_{00} M_{ab} \wedge P^a \wedge P^b + M_{b0}
\wedge P^b \wedge P_0)
\eqno (4.10)
$$
It can be checked that the relation (4.6) is valid, i.e the r-matrix
(4.4) satisfies the modified classical YB equation. In such a case by
duality the cobracket (4.3) induces on the dual Poincar\'e group $G=IO(D)$
the following Poisson bracket $(f_i \equiv f_i (G),  i = 1,2)$  $\alpha = 1
... r =\dim G)$ [18,19] .
$$
\{ f_1, f_2 \} = r^{ \alpha\beta} \left(\partial^R_\alpha f_1 \partial
^R_\beta
f_2 - \partial^L_\alpha f_1 \partial _{\beta}^L f_2 \right)
\eqno (4.11)
$$
where $\partial^R_\alpha$ (resp. $\partial^L_\alpha)$ are the right
(resp. left) invariant vector fields on $G$. If we choose for $G=ISO(D)$
the
$(D+1)$-dimensional matrix representation
$$
t = t_{ij} = {R \ v \choose 0 \ 1};\,\,\,t_{AB} = R_{AB};\,\,\,t_{AD} = v_A;
\,\,\,t_{DD} =1
\eqno (4.12)
$$
where i,j = 0...D ,  $R= (R_{AB})\, (A,B=0...D-1)$ belongs to the fundamental
representation
of a real form of $SO(D, \cal C)$ with the metric $g_{AB}$, i.e.
$R_{AB} R_{CD}g_{BD} = g_{AC}$ and
$v = (v_A)\, \epsilon\, R^D$ describe the translations. The right and left-
invariant vector fields are described by the relations
$< \omega^{R\alpha},\partial^R_\beta > =
< \omega^{L\alpha}, \partial ^L_\beta > = \delta_{\alpha\beta}$
where
$$
\eqalign{\omega^R &= t^{-1} dt = \omega^{R\alpha} e_\alpha \cr
\omega^L &= dtt^{-1} = \omega^{L\alpha} e_\alpha }
\eqno (4.13)
$$
\noindent
Choosing in (4.11) for $f_1, f_2$ the group elements of $G=ISO(D)$
one obtains the following quadratic relations:
$$
\eqalign {\{ t_{im}, t_{jn} \}&= {1 \over \kappa} (t_{ik}(M_{a0})_{km}
\wedge t_{jl} (P^a)_{ln} -(M_{a0})_{ik} t_{km} \wedge (P^a)_{jl} t_{ln}) }
\eqno (4.14)
$$
where $(M_{a0})_{ij} = \delta_{ai} \delta_{0j} + \delta_{aj} \delta_{0i}$
and $(P^a)_{ij} = \delta_{ia}\delta_{jD+1}$ describe the $(D+1)(D+1)$
matrix realizations of $G = IO(D)$, determining the fundamental
matrix realization of the r-matrix (4.4).

The "canonical" Poisson brackets (4.14) on the D-dimensional Poincar\'e
group are compatible with the standard comultiplication
$$
\Delta (t_{im}) = t_{ij} \otimes t_{jm}
\eqno (4.15)
$$
If we perform the standard quantizations of the Poisson brackets (4.14),
by replacing
$$
\{t, t' \} \rightarrow {1\over i} [\hat t, \hat t']
\eqno (4.16)
$$
one obtains the following set of commutation relations:
$$
[\hat R_{AB}, \hat R_{CD}] = 0
\eqno (4.17a)
$$
$$
[\hat R_{AB}, \hat v_C] = {i\over \kappa} \{ (\hat R_{A0}-\delta_{A0})
\hat R_{BC} + g_{AC} (\hat R_{0B} - \delta_{0B} \}
\eqno (4.17b)
$$
$$
[\hat v_A, \hat v_B] = {i \over \kappa} (\hat v_A g_{B0} - \hat v_B g_{A0})
\eqno (4.17c)
$$
with the coproduct (4.15) valid for the non-commutative parameters
$\hat R_{AB}, \hat v_A$ (we recall that $t_{jD} = \hat t_{jD} =
\delta_{jD})$.

It should be stressed that the quantization procedure given in this
paragraph
takes into consideration only the linear terms in the deformation
parameter ${1\over \kappa}$, and consequently the commutators (4.17)
can be expressed by the standard relation $R(t \otimes t)=\tau
(t\otimes t) R$ where $R=\exp ir $. It has been shown in [20] that for
$D=2$ such a quantization is related by a nonlinear transformations of the
parameters to the full quantization, obtained by dualizing the $D=2 \,\,\,
\kappa$-Poincar\'e-algebra (see also [21]). For $D>2$ the relation between the
quantization
obtained from the Poisson bracket (4.11) and the complete one, by
dualizing
the full $\kappa$-Poincar\'e algebra, is under investigation.

\bigskip
\noindent
{\bf 5. Conclusions.}

In the present paper, we described the  quantum
$\kappa$-Poincar\'e-algebra
in any dimension and a version of quantum $\kappa$-Poincar\'e
group obtained by the quantization of the quadratic $r$-matrix Poisson
brackets. We would like finally to mention that :\hfill\break
\medskip
\noindent
i) The universal R-matrix is known only for the case $D=3$ [6] which is
the only dimension (for $D \geq 2)$  for which the quantum de-Sitter
contraction limit of the universal R-matrix for $U_q (o(D+1))$ has provided
after suitable renormalization the
universal R-matrix for $U_q(io(D))$ .
\medskip
\noindent
ii) In order to apply the $\kappa$-Poincar\'e algebra to the description
of the deformed D-dimensional space-time symmetries one should
introduce
the space-time coordinates $x_A$. Two possible ways of introducing
space-time
coordinates can be proposed:
\medskip
\noindent
1) by introducing the commuting space-time coordinates via standard Fourier
transform of the commuting momentum variables.
\medskip
\noindent
2) by considering the non commutative coordinates satisfying the relations
(4.17c), i.e.
$$
[{\hat x}_A, {\hat x}_B] = {i\over \kappa} ({\hat x}_{A} g_{B0} -
{\hat x}_B g_{A0})
\eqno (5.1)
$$

If we introduce the translated space-time coordinates
${\hat x}'_A = {\hat x}_A \oplus {\hat v}_A$, the relations (5.1) are
preserved provided $[{\hat x}_A, {\hat x}_B] = 0$.

It is interesting to develop the $\kappa$-Poincar\'e covariant differential
calculus on the non-commutative coordinates (5.1).
\medskip
\noindent
iii)  Recently the $\kappa$-deformation of the $D=4$ super Poincar\'e
algebra has been obtained by suitable contraction of the quantum
super-de-Sitter algebra $U_q (Osp(1;4)) [22]$. It is known that
(for $3\leq D\leq 10)$ the super-de-Sitter algebra exists only for
$D=3,4,6$ and $10$. It would be interesting to show using e.g. the
algebraic methods [6,7] that the quantum $\kappa$-Poincar'
superalgebra exists for all D-also for those for which the contraction
method from quantum super-de-Sitter algebra cannot be applied.
\bigskip
\noindent
{\bf Acknowledgements:} The first author (J.L.) would like to thank the
University of Geneva for its hospitality and the Swiss National Science
Foundation for its financial support. H.R. thanks H. Bacry, R. Duval and
D. Kastler for discussions,  the CNRS Luminy for its
hospitality and the Universit' de Provence (Aix-Marseille I) for its
financial support.

\bigskip

\noindent{\bf References:}

\item{[1]} V.G. Drinfeld, Quantum Groups, Proc. Intern. Congress of
Mathematicians (Berkeley, CA, USA), vol I, p. 789 (1986).

\item{[2]} M. Jimbo, Lett. Math. Phys. \underbar{10} (1985) 63; \underbar{11}
(1986) 247.

\item{[3]} E. Celeghini, R. Giachetti, E. Sorace and M. Tarlini, J. Math.
Phys. \underbar{31} (1990) 2548; 32 (1991) 1155; \underbar{32} (1991) 1159 and
"Contractions of quantum groups" in Quantum Groups, Lecture Notes in
Mathematics No 1510, 221, (Springer-Verlag, 1992).

\item{[4]} J. Lukierski, A. Nowicki, H. Ruegg and V. N. Tolstoy, Phys. Lett.
\underbar{B264} (1991) 331. S. Giller, P. Kosinski, M. Majewski, P.
Maslanka and J. Kunz, Phys. Lett. \underbar{B286} (1992) 57

\item{[5]} J. Lukierski, A. Nowicki and H. Ruegg, Phys. Lett. \underbar{B293}
(1992) 344.

\item{[6]} H. Bacry, "Which Deformations of the Poincar\'e Algebra ?" CNRS
Luminy Preprint, CPT-93/P. 2880.

\item{[7]} J. Lukierski, A. Nowicki and H. Ruegg, "Quantum Deformation of
Nonsemisimple algebras: the Example of $D=4$ Inhomogeneous
Rotations", University of Geneva, preprint UGVA-DPT 1993/3-812.

\item{[8]} S. Majid, J. Math. Phys. \underbar{34} (1992) 2045.

\item{[9]} S. Woronowicz, private communications.

\item{[10]} M. Chaichian and A. Demichev, Phys. Lett. \underbar{B304} (1993)
220 and Helsinki Univ. Preprint HU-TFT-93-24.

\item{[11]} J. Lukierski, H. Ruegg and W. R\"uhl, "From $\kappa$-Poincar\'e
 Algebra to $\kappa$-Lorentz Quasigroup: A Deformation of
Relativistic Symmetry",  Physics Letters B313 (1993) 357-366.

\item{[12]} G. Domokos and S. Kovesi-Domokos, "Astrophysical Limit on the
Deformation of the Poincar\'e Group", preprint JHK-TIPAC-920027/REV. May
1993.

\item{[13]} L. C. Biedenharn, B. Mueller and M. Tarlini, "The Dirac-Coulomb
Problems for the $\kappa$-Poincar\'e Quantum Group", Universities of Texas
at Austin, Duke and Firenze, Preprint 1993.

\item{[14]} H. Ruegg, "q-Deformation of Semisimple and Non-Semisimple Lie
Algebras", in "Integrable Systems, Quantum
Groups and Quantum Field Theories", Ibort and Rodriguez ed., Kluwer
Academic Publishers, Dordrecht (1993) pp. 45-81.

\item{[15]} P. Maslanka, "Deformation Map and Hermitean Representations of
$\kappa$-Poincar\'e Algebra", Preprint Lodz University 1/93, IM UL.

\item{[16]} S. Zakrzewski, "Quantum Poincar\'e Group Related to
$\kappa$-Poincar\'e
Algebra", Warsaw University Preprint, 1993, submitted to Lett. Math.
Phys.

\item{[17]} I.M. Gelfand and I. Ya. Dorfman, Funct. Anal. i Prilozhen
\underbar{16}, no 4, p 1 (1982) (in Russian).

\item{[18]} V. G. Drinfeld, Dokl, Akad. Nauk. SSSR \underbar{268}, (1983) 285.

\item{[19]} M.A. Semenov-Tian-Shansky, Publ. RIMS Kyoto Univ. \underbar{21}
(1985) 1237.

\item{[20]} P. Maslanka, "The Two-Dimensional Quantum Euclidean Group" Lodz
University Preprint 2/93, IM UL.

\item{[21]} A. Balesteros, E. Celeghini, R. Giachetti, E. Sorace and M.
Tarlini,
"An R-Matrix approach to the Quantization of the Euclidean Group E(2)",
University of Firenze, Preprint DFF 182/1/93.

\item{[22]} J. Lukierski, A. Nowicki and J. Sobczyk, University of Wroclaw
preprint ITP UWr 890/93, March 1993; J. of Phys. A., in press.

\bye